\documentclass[12pt]{article}

\usepackage{epsfig}

\newcommand{\lwig}{\mbox{\,\raisebox{.3ex}
    {$<$}$\!\!\!\!\!$\raisebox{-.9ex}{$\sim$}\,}}
\newcommand{\gwig}{\mbox{\,\raisebox{.3ex}
    {$>$}$\!\!\!\!\!$\raisebox{-.9ex}{$\sim$}}\,}
\newcommand{\xbj}{x_{\rm Bj}}

\newcommand{\xpr}{{x^\prime}}
\def\np#1#2#3{Nucl.\ Phys.\ {\bf B#1}, #2 (19#3)}
\def\pl#1#2#3{Phys.\ Lett.\ {\bf #1B}, #2 (19#3)}
\def\pr#1#2#3{Phys.\ Rev.\ D {\bf #1}, #2 (19#3)}

\def\prl#1#2#3{Phys.\ Rev.\ Lett.\ {\bf #1}, #2 (19#3)}

\def\zp#1#2#3{Zeit.\ Phys.\ {\bf C#1}, #2 (19#3)}
\date{}

\begin{document}
\title{{\normalsize\rightline{DESY 97-115}\rightline{hep-ph/9706399}} 
  \vskip 1cm 
  \bf Instanton Phenomenology at HERA\thanks{Talk presented at the 5th
    International Workshop on Deep-Inelastic Scattering and QCD (DIS\,97),
    Chicago, April 1997; to be published in the Proceedings (AIP).}}
\vspace{2cm}
\author{A. Ringwald and F. Schrempp\\[0.5cm]
{\it\small DESY, Notkestr. 85, D-22603 Hamburg, Germany}}

\begin{titlepage} 
\maketitle

\begin{abstract}
This talk describes the physics input of QCDINS, a Monte Carlo event 
generator for QCD-instanton induced scattering processes in 
deep-inelastic scattering. 
\end{abstract}
\thispagestyle{empty}
\end{titlepage}
\newpage \setcounter{page}{2}

Hard scattering processes in strong interactions are                   
successfully described by perturbative QCD. 
However, perturbation theory does not exhaust all
possible hard scattering processes: Instantons \cite{bpst}, non-perturbative
fluctuations of the gluon fields, induce hard processes which are
absent in perturbative QCD. Deep-inelastic scattering (DIS) 
at HERA offers a unique window to detect these processes through their 
characteristic multi-particle final-state signature \cite{rs}. 
A Monte-Carlo generator for instanton-induced events in DIS, 
QCDINS, interfaced with HERWIG, has been developed \cite{grs} 
which enabled the H1 Collaboration to place first experimental upper
limits on the cross-section \cite{h1_limits} and which allows for the
elaboration of dedicated search strategies \cite{ggmrs}. 
The purpose of this talk is to outline the basic physics input of QCDINS 
and its built-in features, characteristic for the underlying instanton 
mechanism. Further details about the theoretical background \cite{mrs_dis97}
and ongoing experimental searches \cite{ck} appear elsewhere in these
proceedings.

In order to appreciate the notion of instanton-induced scattering
processes, let us recall that scattering amplitudes are derived via analytic
continuation and LSZ reduction 
from Euclidean Green's functions, which in turn can be represented by a
path integral,
\begin{equation}\label{eq:pathintegral}
\frac{1}{Z}\, 
\int [dA][d\psi][d\overline{\psi}] A_{\mu}(x_{1})\ldots \psi (x_{i})\ldots
             \overline{\psi}(x_{n})
        \exp\{-S\,[A,\psi,\overline{\psi}]\} .
\end{equation}
 
\begin{figure}[ht]
\begin{center}
\epsfig{file=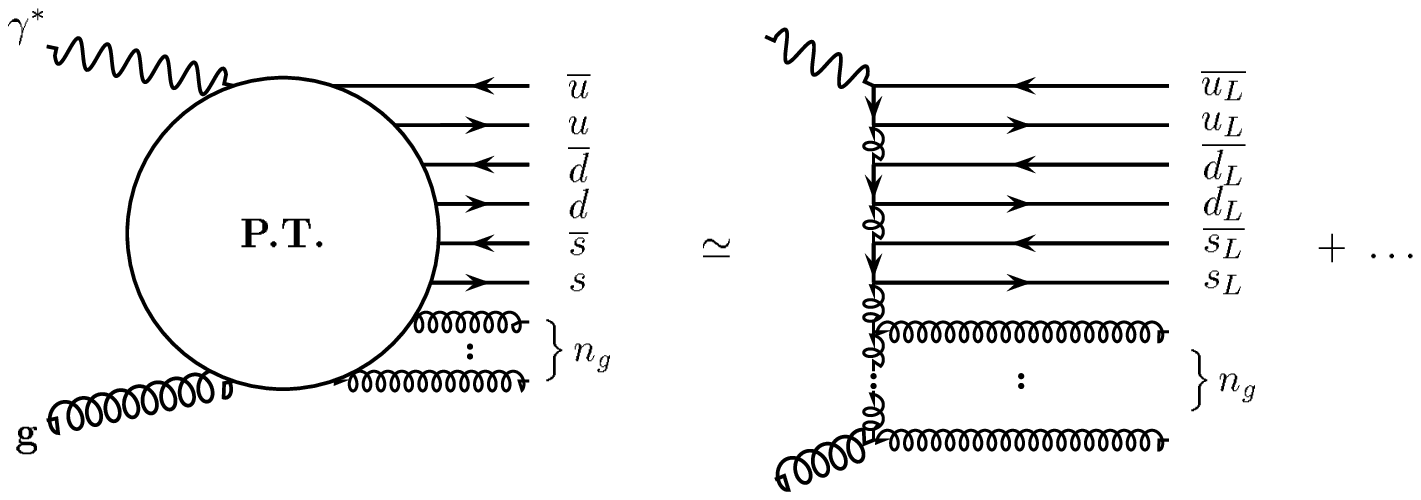,%
width=6.8cm}\hfill
\epsfig{file=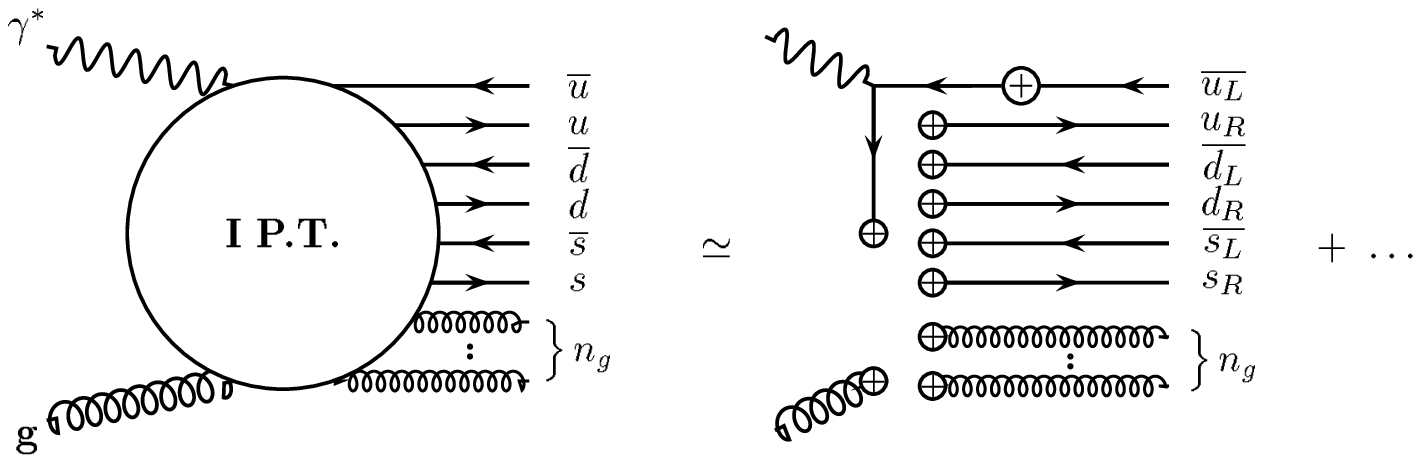,%
width=6.8cm}
\caption[]{Tree amplitudes for $\gamma^{\ast}g$ 
      scattering processes, in which all light quark 
      flavors are produced. Left: Usual perturbation theory.
      Right: Instanton-perturbation theory. Lines ending with
      a blob stand for classical right-handed quark $(\psi_{R}^{(I)})$ and
      anti-quark $(\overline{\psi_{L}}^{(I)})$ solutions; line with
      a blob in the middle denotes the quark propagator in the $I$-background;
      curly lines ending with a blob stand for classical instanton
      gluon fields $(A_{\mu}^{(I)})$.
}
\label{fig:comp}
\end{center}
\end{figure}

The perturbative scattering amplitudes are obtained from an 
expansion of Eq.~(\ref{eq:pathintegral})
about the {\it perturbative-vacuum solution}, i.e. vanishing gluon fields,
 $A_{\mu}^{(0)}=0$, and vanishing quark fields, $\psi^{(0)}=
\bar{\psi}^{(0)} =0$, with vanishing Euclidean action $S^{(0)}=0$.
This expansion can be summarized by the familiar Feynman rules which construct
the perturbative scattering amplitudes in terms of propagators and
vertices as a power-series in the strong coupling $\alpha_{s}$ 
(see Fig.~\ref{fig:comp} (left)). 

The {\it instanton $A^{(I)}_{\mu}(x)$} is a {\it non-trivial solution}
of the Euclidean gluon-field equations and thus a non-trivial local minimum 
of the Euclidean action with $S^{(I)}=2\pi /\alpha_{s}$. Instanton-induced
scattering amplitudes, being derived from an expansion of 
Eq.~(\ref{eq:pathintegral}) about the instanton, can be constructed 
according to modified Feynman rules, which involve, in addition to the 
propagators (in the $I$-background) and vertices (in the $I$-background) 
also the classical fields (see Fig.~\ref{fig:comp} (right)). Instanton-induced
scattering amplitudes are always exponentially suppressed at weak coupling, 
$\propto\exp\{-{ 2\pi /\alpha_{s}}\}$.

In QCD with massless quarks, usual perturbation theory
and instanton perturbation theory describe two distinct classes
of processes: In usual perturbation theory, quarks are coupled to gauge
fields only via vector couplings, which {\it conserve chirality} ($Q_{5}$).
Thus quarks always appear in pairs with net chirality zero (see
Fig.~\ref{fig:comp} (left)) and chirality is conserved to all orders.
In instanton-perturbation theory, on the other hand, only scattering
amplitudes for processes which {\it violate chirality} by  $\triangle 
Q_{5}=2n_{f}$ units receive non-vanishing contributions \cite{th}. 
This is due to the fact that the classical quark solutions always appear 
in pairs with net chirality two (see Fig.~\ref{fig:comp} (right)).

The Monte-Carlo simulation of $I$-induced events proceeds in
three steps. First, quasi-free partons are produced by QCDINS with the
distributions prescribed by the hard process matrix elements. Next, 
these primary partons give rise to parton showers, as described
by HERWIG. Finally, the showers are converted into hadrons, again
within HERWIG. 

\begin{figure}
\begin{center}
\epsfig{file=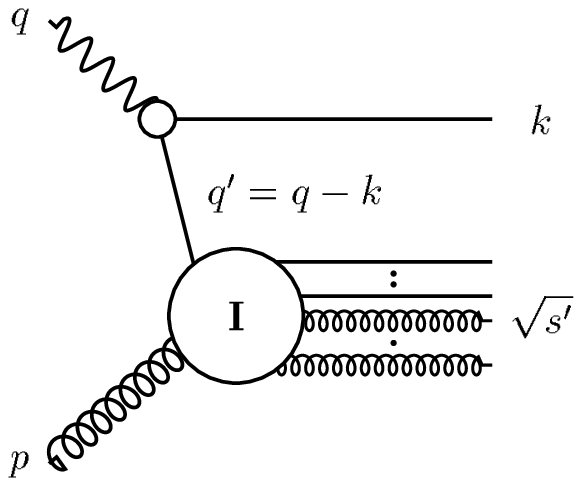,height=4cm}\hfill
\epsfig{file=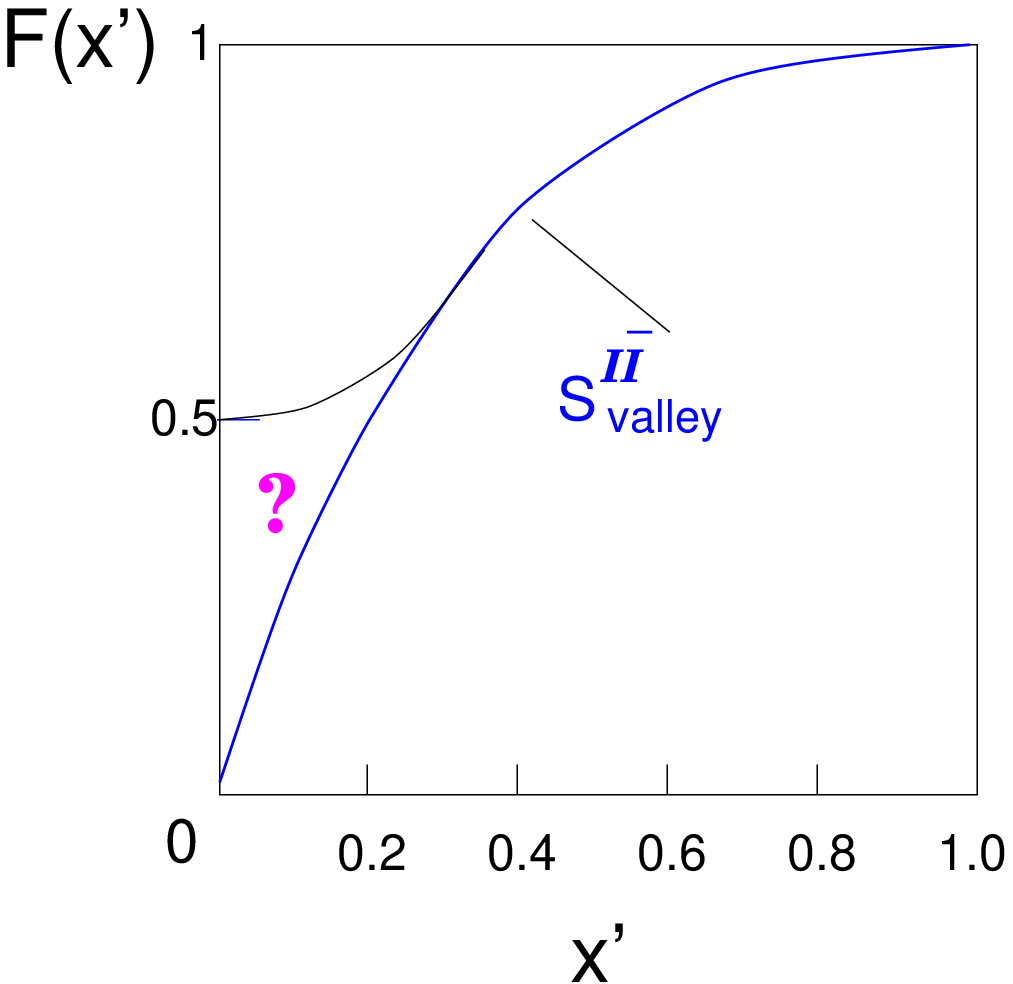,height=4cm}\hfill
\epsfig{file=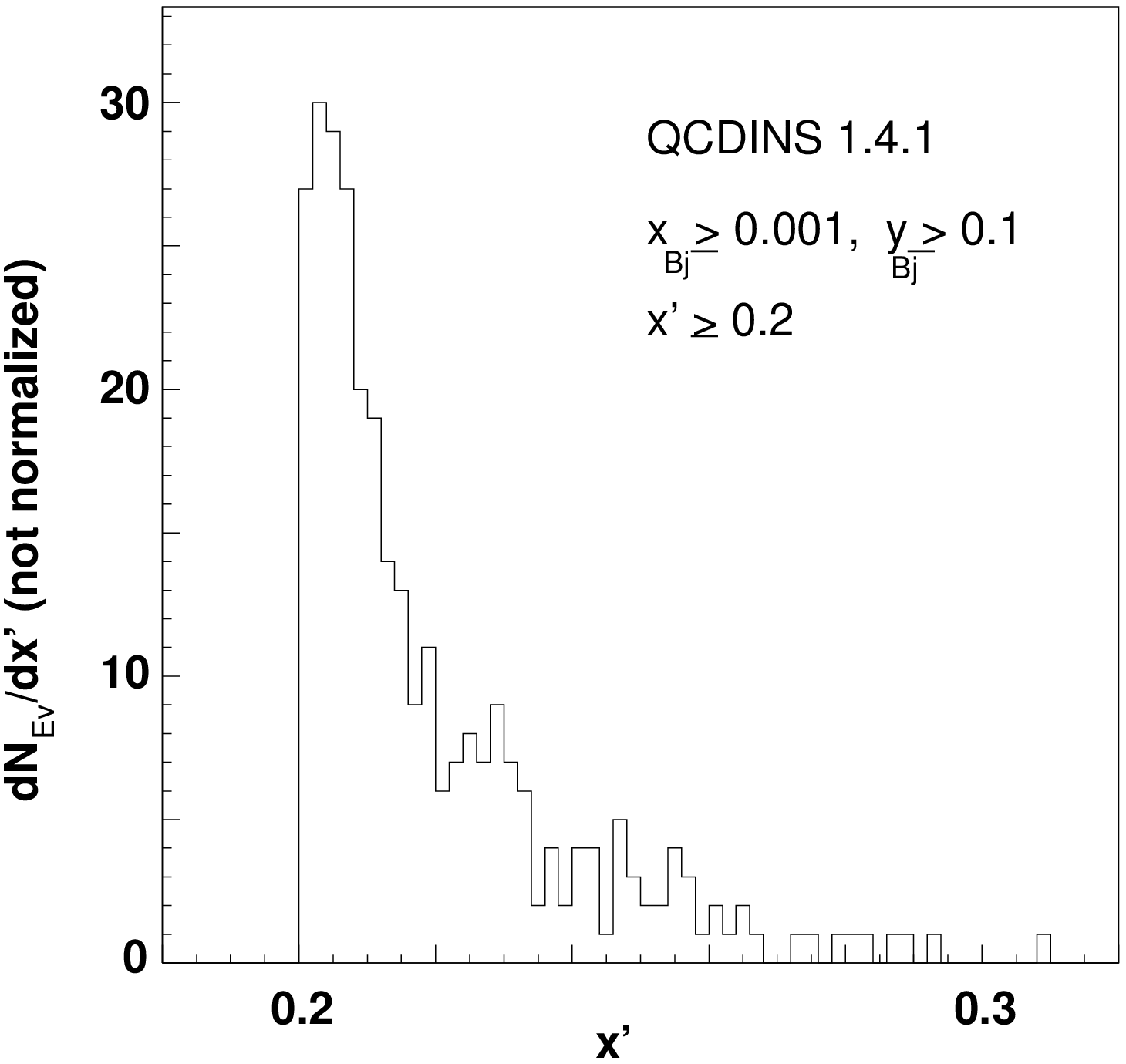,height=4cm}
\caption[]{Left: Momentum-space structure of $I$-induced partonic processes in 
$\gamma^{\ast}g$ scattering. Middle: The holy-grail function $F(\xpr)$.
Right: $x^\prime$ distribution from QCDINS.}
\label{fig:me}
\end{center}
\end{figure}

The momentum-space structure of $I$-induced hard processes in 
$\gamma^{\ast}g$ scattering is shown in Fig.~\ref{fig:me} (left). 
As is already suggested by the form of the
leading-order matrix elements in Fig.~\ref{fig:comp} (right), the amplitudes
factorize into a product of an effective $\gamma^{\ast}qq^{\ast}$ 
vertex \cite{mrs}, denoted by a blob in Fig.~\ref{fig:me} (left), times matrix 
elements for $I$-induced partonic subprocesses, 
$q^{\ast}g\to (2n_{f}-1)\,q+n_{g}g$, denoted by a blob with an index ``$I$''. 
It turns out that the most important kinematical variables determining
the final state properties of $I$-induced events are the 
virtuality of the off-shell quark 
$q^{\ast}$, $Q^{\prime 2}\equiv -q^{\prime 2}\geq 0$, 
and the Bjorken scaling variable of the $q^{\ast}g$ subprocess, 
$x^{\prime}\equiv Q^{\prime 2}/(2p\cdot q^{\prime})$. Therefore let us 
discuss their distributions first. 
   
To this end, we square the amplitudes in Fig.~\ref{fig:me} (left) and sum 
over all final state partons to construct the $I$-induced 
hard-scattering partonic total cross-section or structure function. 
It can be shown \cite{mrs} that the initial state collinear singularities, 
occuring when $Q^{\prime 2}\to 0$, can be consistently absorbed into the parton
distribution functions $f_{k}$, such that the $I$-induced contribution to the 
nucleon-structure function is obtained in the familiar convolution form,    
\begin{equation}\label{eq:conv}
  F_2^{{ (I)}} (\xbj ,Q^2)  =
   \sum_{k} \int\limits_{\xbj}^1 \frac{dx}{x}\, 
   f_{k}\left( \frac{\xbj}{x},\mu_{f}^{2},\mu^{2}\right)  \,\frac{\xbj}{x}\
   {\mathcal C}^{{ (I)}}_{ 2\, {k}} {
   \left( x ,\frac{Q^2}{\mu^{2}},
   \frac{\mu_{f}^{2}}{\mu^{2}},\alpha_{s}(\mu)\right)} \, , 
\end{equation}
where $\mu (\mu_{f})$ denotes the renormalization (factorization) scale. 
The $I$-contribution to the dominating gluon-coefficient function 
${\mathcal C}_{2\,{g}}^{{(I)}}$ in turn, has for large $Q^{2}$ 
the anticipated momentum-space structure \cite{rs,mrs},
\begin{eqnarray}\label{eq:coeff}
\lefteqn{   
{\mathcal C}_{2\,{g}}^{{(I)}} 
\left( x ,\frac{Q^2}{\mu^{2}},
   \frac{\mu_{f}^{2}}{\mu^{2}},\alpha_{s}(\mu)\right)  
   \simeq x\sum_{q}e_{ q}^2\,\times 
} \\ \nonumber &&  
   \int\limits_x^1 \frac{dx^\prime}{x^\prime} 
   \int\limits_{\mu_{f}^{2}}^{Q^2\frac{x^\prime}{x}}
   dQ^{\prime 2}\, \frac{3}{16\,\pi^3}\,\frac{x}{x^\prime}
\left(1+\frac{1}{x}-\frac{1}{x^{\prime}}-\frac{Q^{\prime 2}}{Q^{2}}\right)
  \sum_{n_{g}}\sigma^{(I)}_{{q}^\ast { g};\,{ n_{g}}} 
   \left( x^\prime, Q^{\prime 2},\alpha_{s}(\mu)\right) .
\end{eqnarray}
The essential instanton dynamics and in particular most of the dependence on
$\xpr$ and $Q^{\prime}$ is encoded in the $I$-subprocess total
cross section,   
\begin{eqnarray}\label{eq:cross-section}
\sum_{n_{g}}\sigma^{(I)}_{{q}^\ast { g};\,{ n_{g}}}
(\xpr ,Q^{\prime 2})
 \simeq\frac{\Sigma (\xpr )}{Q^{\prime 2}}\ 
\left( \frac{4\pi}{\alpha_s (\mu (Q^\prime ) )}\right)^{21/2}
\ \exp\left[-\frac{4\pi}{\alpha_s (\mu (Q^\prime ) )}
\,{ F(\xpr )}\right]\ 
,
\end{eqnarray}
where the functions $\Sigma (\xpr )$ and $F(\xpr )$ (see Fig.~\ref{fig:me} 
(middle)) are known \cite{bb,rs} for 
$\xpr \geq x^{\prime}_{\rm min} \simeq 0.2-0.3$.

We see from Eq.~(\ref{eq:cross-section}) that the summation over the gluon
emission has modified the exponential suppression factor by the so-called
{``holy-grail'' function} $ F(\xpr )$. This exponentiation is due to the fact 
that every gluon emission brings in a factor of 
$A^{(I)}_{\mu}\sim 1/\sqrt{\alpha_{s}}$ in the exclusive amplitudes 
(see Fig.~\ref{fig:comp} (right)), such that \cite{r}   
$\sigma^{(I)}_{{q}^\ast { g};\,{n_{g}}} \propto   (1/n_{g}!)
(1/\alpha_{s})^{ n_{g}} 
                 \exp\{-4\pi/\alpha_{s}\}$.

The holy-grail function $F(\xpr )$ is continuously decreasing from 
$1$ at $\xpr =1$ (low $q^{\ast}g$ c.m. energy) to $1/2$ at 
$\xpr \simeq 0.2$ (see Fig.~\ref{fig:me} (middle)).  The total
cross-section is correpondingly exponentially growing with decreasing
$\xpr$. This is clearly seen in the $\xpr$ distribution taken from
the Monte-Carlo simulation (see Fig.~\ref{fig:me} (right)). 

\begin{figure}[ht]
\begin{center}
\epsfig{file=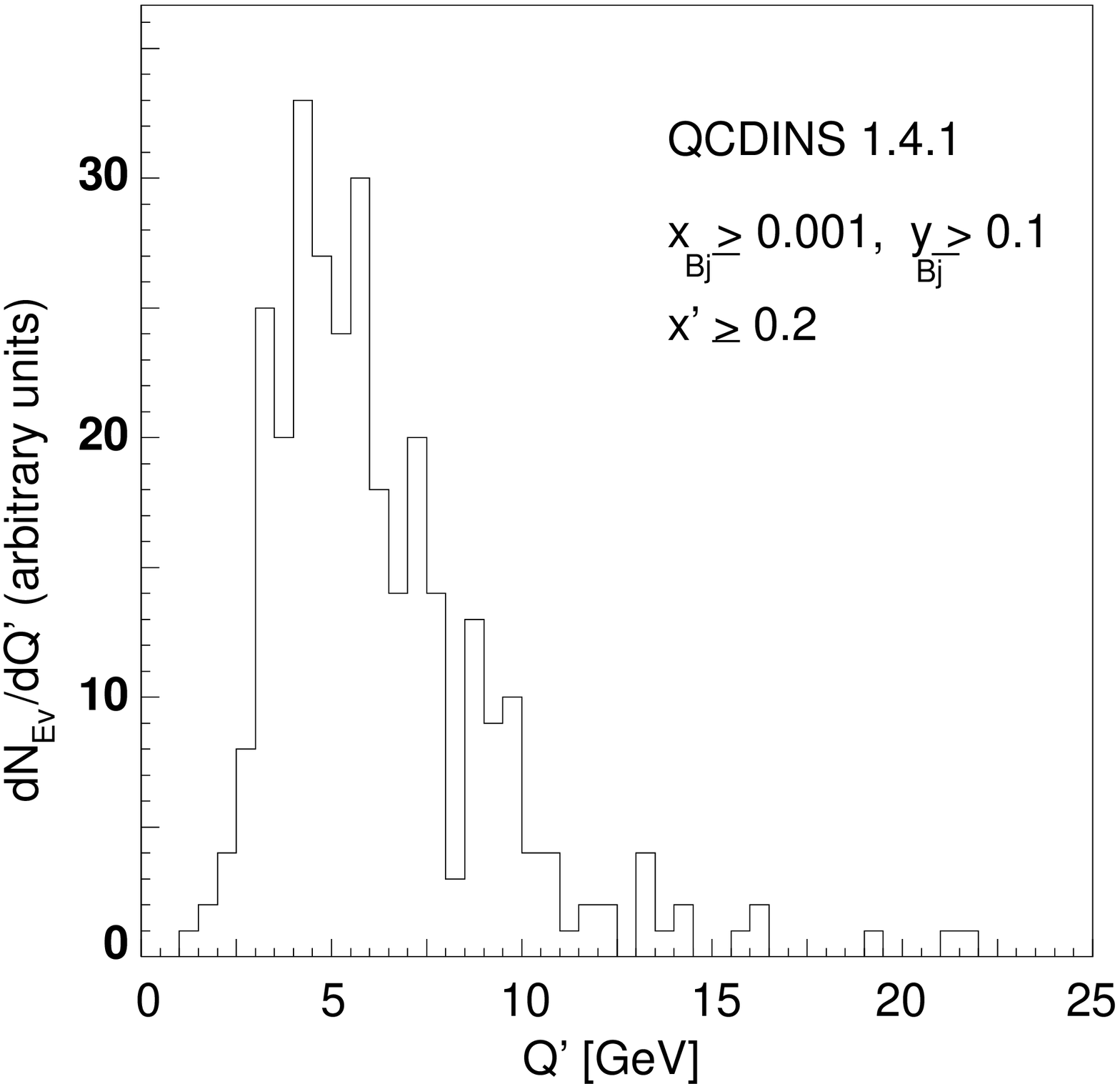,height=4.6cm}\hfill
\epsfig{file=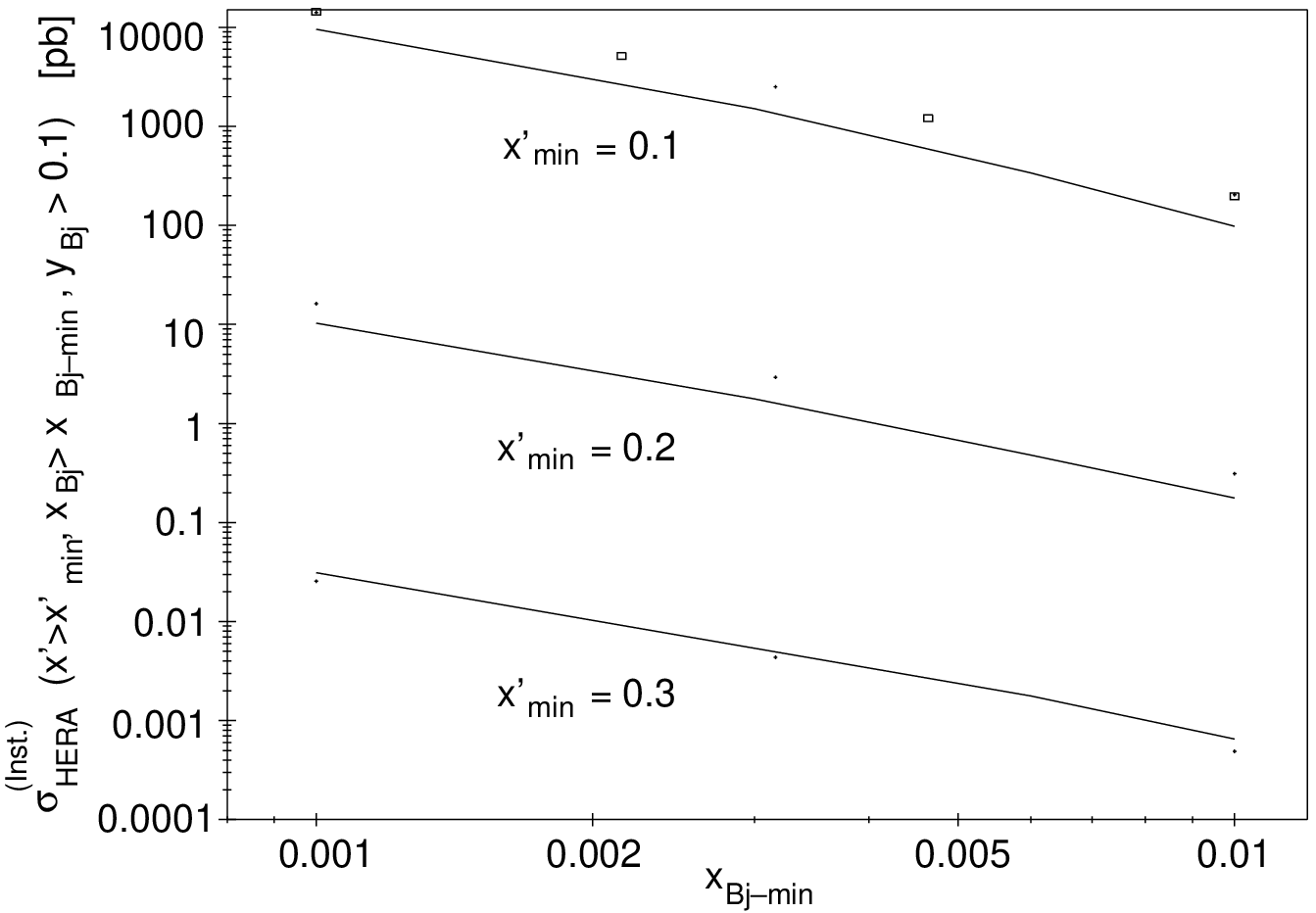,height=4.6cm}
\caption[]{Left: $Q^{\prime}$ distribution. Right: $I$-induced total 
$eP$ cross-section for HERA (preliminary) with 
various cuts as indicated. Lines: Analytical calculation. Points: Monte-Carlo 
simulation with QCDINS 1.4.1.}
\label{fig:i_dist}
\end{center}
\end{figure}

Thanks to the chosen renormalization scale, 
$\mu (Q^\prime)= Q^{\prime } 
\alpha_{s}(\mu(Q^\prime ))/(4\pi)$, the $I$-subprocess cross-section
(\ref{eq:cross-section}), as a function of the $q^{\ast}$ virtuality 
$Q^{\prime}$, has a peak structure which is clearly reflected
by the $Q^{\prime}$ distribution in Fig.~\ref{fig:i_dist} (left). The peak
is at around 5 GeV, which is gratifying since it means that even without 
a lower $Q^{\prime}$ cut\footnote{A minimum $Q^{\prime}$ can 
be enforced by requiring the transverse momentum $k_{T}$ of the current-quark 
jet (see Fig.~\ref{fig:me} (left)) to be large. A lower $\xpr$-cut can be
implemented, for example, by requiring 
$x\equiv Q^{2}/(Q^{2}+s)\gwig\, x^{\prime}_{\rm min}$, 
where $s$ is the $\gamma^\ast g$ c.m. energy. \label{foot:cuts}} most
of the events generated are hard enough in order to justify
instanton-perturbation theory \cite{mrs}.     

In Fig.~\ref{fig:i_dist} (right) we present the resulting $I$-induced
total $eP$ cross-section for HERA, subject to the following cuts:
{\it i)} $x_{\rm Bj}\geq x_{\rm Bj\ min},\, y_{\rm Bj}\geq 0.1$ in 
         Eq.~(\ref{eq:conv});
{\it ii)} $\xpr\geq x^{\prime}_{\rm min}$ in the integration of 
         Eq.~(\ref{eq:coeff}).
No cut on $Q^{\prime}$ has been imposed. It is important to note that 
lower limits on $\xpr$ and $Q^{\prime}$ can be enforced experimentally by
cuts on final state momenta$^{\ref{foot:cuts}}$. 
The points in 
Fig.~\ref{fig:i_dist} (right) have been taken from the Monte-Carlo simulation 
(QCDINS 1.4.1). As a check of the Monte-Carlo, we have also analytically
integrated Eqs.~(\ref{eq:conv}) and (\ref{eq:coeff}). The resulting 
cross-section (lines in Fig.~\ref{fig:i_dist} (right)) nicely agrees with 
the Monte-Carlo result. We refrain from going to even smaller
$x_{\rm Bj}$, say $10^{-4}$, since in this case 
$Q^{2}=x_{\rm Bj}\,y_{\rm Bj}\,S$ would be only of order $1$ GeV$^{2}$. It is
not clear whether in this case the corrections to Eq.~(\ref{eq:coeff}), 
which are not built into QCDINS, can be neglected.  
     
As a benchmark for searches for $I$-induced events at HERA 
\cite{h1_limits,ck}, let us compare the $I$-induced HERA cross-section from 
Fig.~\ref{fig:i_dist} (right) with the normal DIS HERA cross-section, 
$\sigma_{\rm HERA}^{{\rm (nDIS)}}
              (x_{\rm Bj}>10^{-3},y_{\rm Bj}>0.1)\simeq 15\ {\rm nb}$.  
We see that the fraction of $I$-induced events at HERA,
$f^{(I)}\equiv \sigma_{\rm HERA}^{{(I)}}/
                     \sigma_{\rm HERA}^{{\rm (nDIS)}}$, 
ranges between
\begin{equation}
{ 0.0002\ \%}\ \lwig  f^{(I)} \lwig\ { 0.1\ \%}\, ,
\end{equation}
in the kinematical range 
$\xpr >0.3..0.2,\ x_{\rm Bj}>10^{-3},\ y_{\rm Bj}>0.1$. 
Note that the published upper limits on the fraction of $I$-induced
events placed by the H1 Collaboration \cite{h1_limits} are in the
several percent range. An improvement of these limits by an order of 
magnitude is reported in another talk in this working group \cite{ck}.   

\begin{figure}[ht]
\begin{center}
\epsfig{file=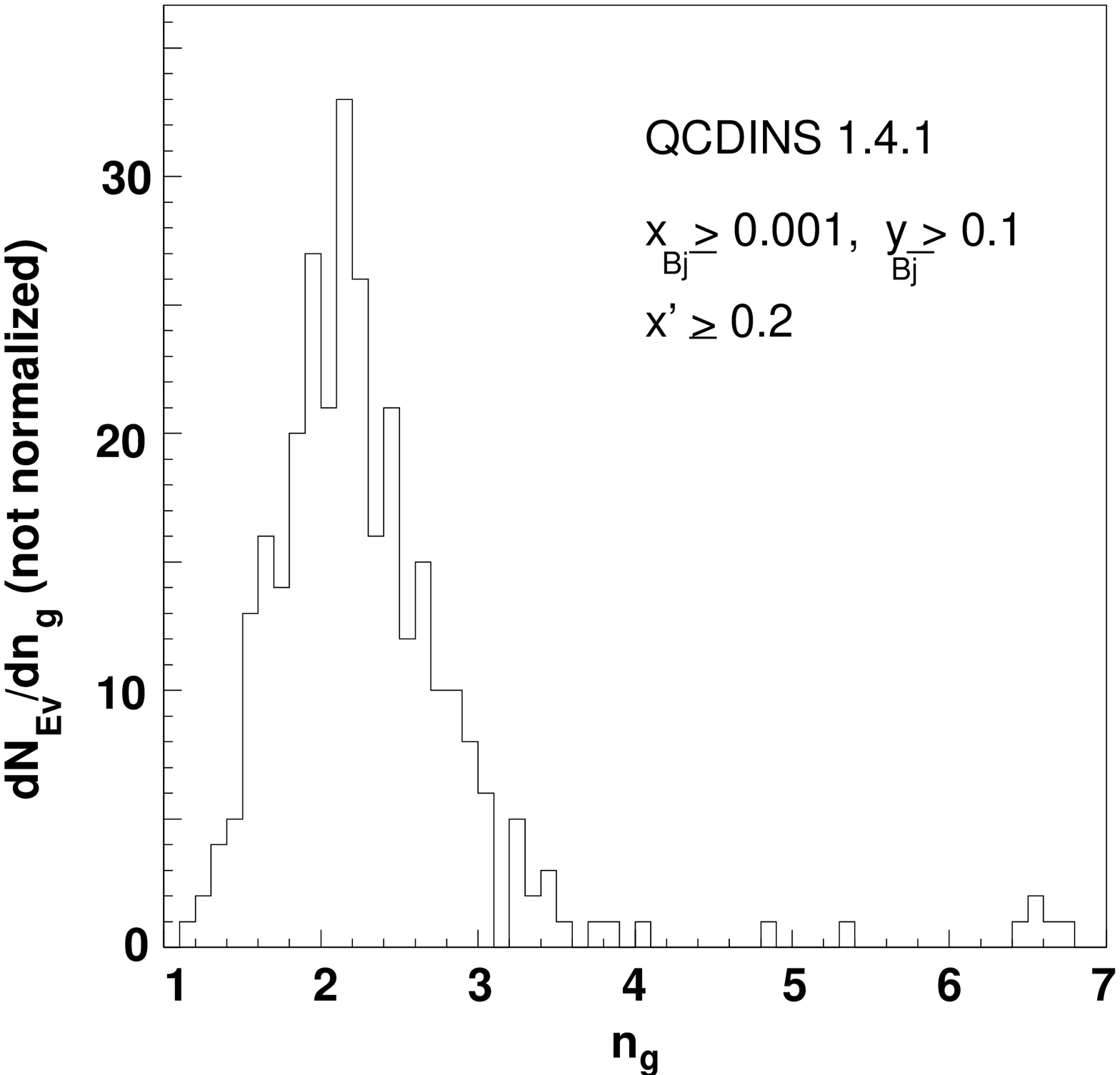,height=5cm}\hfill
\epsfig{file=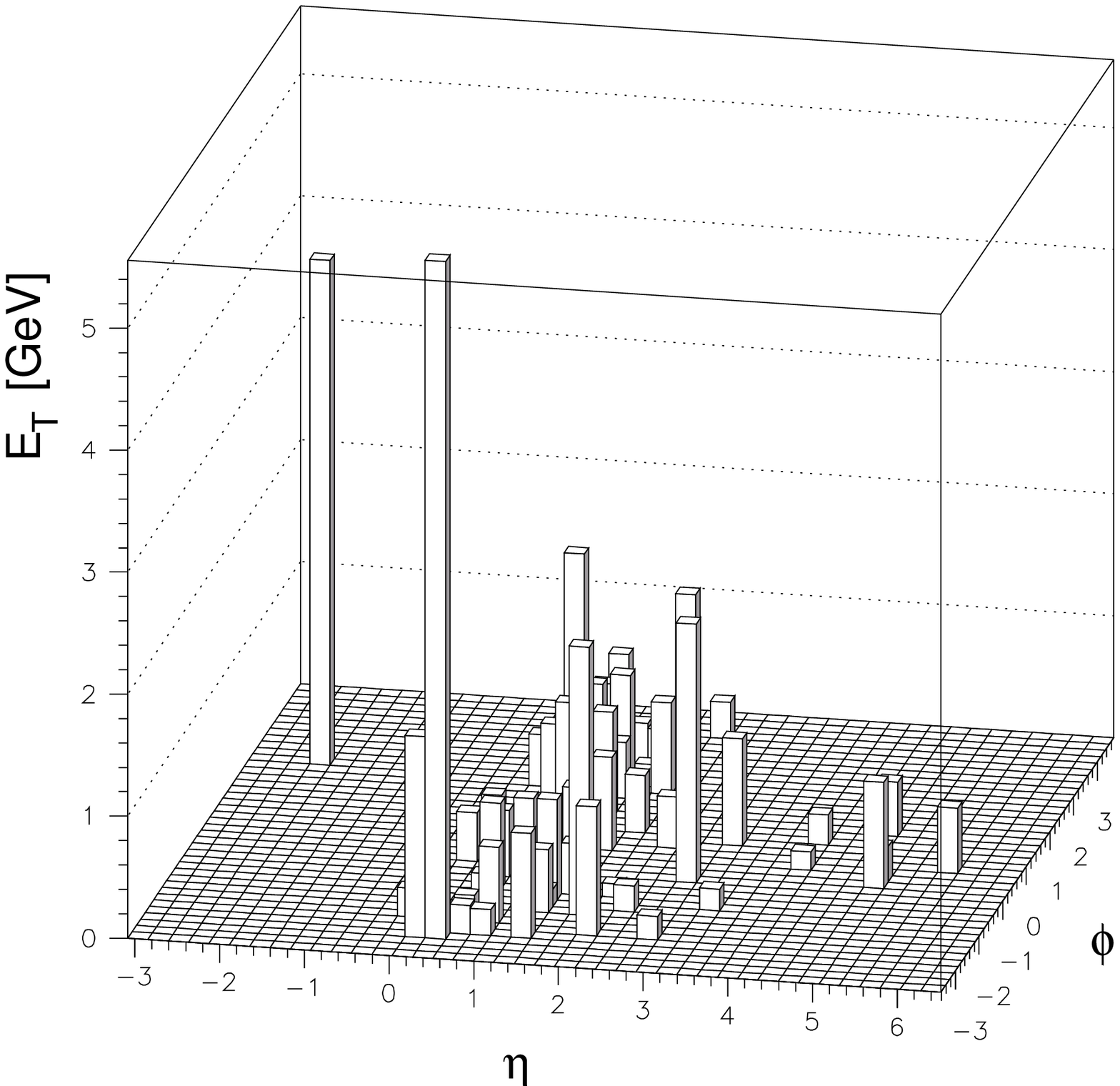,height=5cm}
\caption[]{Left: Gluon multiplicity distribution of $I$-induced events. 
Right: Lego plot of a typical $I$-induced event in the HERA-lab system
at $x_{\rm Bj}=10^{-3}$.}
\label{fig:typical}
\end{center}
\end{figure}

Let us turn now to the final states of $I$-induced events in DIS.
The current quark in Fig.~\ref{fig:me} (left) will give rise, after 
hadronization, to a current-quark jet. The partons from the $I$-subprocesses, 
$q^{\ast}g\to (2n_{f}-1)\,q+n_{g}g$ (see Fig.~\ref{fig:me} (left)), 
on the other hand, are emitted spherically symmetric in the   
$q^{\ast}g$ c.m. system (``$I$-c.m. system''). The gluon multiplicities are 
generated according to a Poisson distribution with mean multiplicity
\begin{equation}
\langle n_g (\xpr ,Q^{\prime})\rangle^{{(I)}}\simeq 
{\frac{2\pi}{\alpha_{s}(\mu(Q^{\prime}))}}\,\xpr (1-\xpr) 
\frac{dF(\xpr)}{d\xpr} ,
\end{equation}
which is of the order of two for $\xpr\simeq 0.2$, $Q^{\prime}\simeq 5$ GeV 
(see Fig.~\ref{fig:typical} (left)). 
The total mean parton multiplicity is large, of the order of ten. 
After hadronization we therefore expect from the $I$-subprocess a
final state structure reminiscent of a decaying fireball: Production
of the order of 20 hadrons, always containing strange and possibly charmed 
mesons, concentrated in a ``band'' at fixed pseudorapidity $\eta$ in
the ($\eta$, azimuth angle $\phi$)-plane. Due to the boost 
from the $I$-c.m. system to the HERA-lab system, the center of the
band is shifted in $\eta$ away from zero, and its half-width is 
of order $\Delta\eta =0.9$, as typical for a spherically symmetric event. 
The total invariant mass of the 
$I$-system, $\sqrt{s^{\prime}}=Q^{\prime}\sqrt{1/\xpr -1}$, 
is expected to be in the 10 GeV range, for $\xpr\simeq 0.2$, 
$Q^{\prime}\simeq 5$ GeV. The lego plot of a typical $I$-induced event
shown in Fig.~\ref{fig:typical} (right) shows that these expectations
(current-quark jet; hadronic ``band'') are actually bourne out from
our Monte-Carlo simulation.

\section*{Acknowledgements}
We would like to thank M. Gibbs and S. Moch for fruitful collaboration
and T. Carli for its help in debugging and improving QCDINS.

\end{document}